\documentclass[12pt]{article}
\usepackage{amssymb,amsmath}
\usepackage{epsfig}
\usepackage{epstopdf}
\usepackage{latexsym}

\numberwithin{equation}{section}

\renewcommand{\theequation}{\arabic{section}.\arabic{equation}}

\def\coeff#1#2{\relax{\textstyle {#1 \over #2}}\displaystyle}
\def\ds{\displaystyle}

\def\cN{{\cal N}}

\newcommand{\be}{\begin{equation}}
\newcommand{\ee}{\end{equation}}
\newcommand{\bea}{\begin{eqnarray}}
\newcommand{\eea}{\end{eqnarray}}

\begin{document}

\begin{titlepage}

\begin{flushright}
\end{flushright}

\bigskip
\bigskip
\centerline{\Large \bf The Nuts and Bolts of Einstein-Maxwell Solutions}
\bigskip
\bigskip
\bigskip
\centerline{{\bf  Nikolay Bobev$^1$ and Cl\'{e}ment
Ruef$\, ^2$ }}
\bigskip
\centerline{$^1$Department of Physics and Astronomy, University of
Southern California,} \centerline{Los Angeles, CA 90089, USA}
\bigskip
\centerline{$^2$Institut de Physique Th\'eorique, } \centerline{CEA
Saclay, 91191 Gif sur Yvette, France}
\bigskip
\centerline{{\rm bobev@usc.edu\,,
~clement.ruef@cea.fr} }
\bigskip \bigskip

\begin{abstract}
We find new non-supersymmetric solutions of five-dimensional ungauged supergravity coupled to two vector multiplets. The solutions are regular, horizonless and have the same asymptotic charges as non-extremal charged black holes.  An essential ingredient in our construction is a four-dimensional Euclidean base which is a solution to Einstein-Maxwell equations. We construct stationary solutions based on the Euclidean dyonic Reissner-Nordstr\"{o}m black hole as well as a six-parameter family with a dyonic Kerr-Newman-NUT base. These solutions can be viewed as compactifications of eleven-dimensional supergravity on a six-torus and we discuss their brane interpretation.  

\end{abstract}

\end{titlepage}

\section{Introduction}

Finding and understanding supersymetric solutions of supergravity theories is a very important task, and significant advances have been achieved in this direction. For example, one of the major result is the classification of all supersymmetric solutions of five-dimensional minimal supergravity \cite{Gauntlett:2002nw}.  To find and classify supersymmetric solutions one is typically utilizing the supersymmetry variations of the fermionic fields, which lead to first order differential equations that are more tractable than the second order equations of motion. Undoubtedly, supersymmetric gravity solutions have very interesting physics, some intriguing mathematical structure and provide a good laboratory for testing new ideas on tractable examples. However, one would ultimately like to construct and understand non-supersymmetric and non-extremal solutions and it is important to have as much exact solutions as possible to gain intuition about their structure and properties.

Of separate, albeit related, interest are asymptotically flat supergravity solutions with no horizons and singularities. Such regular solutions may represent possible microstates for black holes (or black rings) having the same charges and asymptotic structure. This idea, first proposed by Mathur, is the essence of the fuzzball proposal and it has been implemented with a growing success for BPS black holes and black rings with two and three charges, see \cite{Mathur:2005zp} for reviews and further references. In the supersymmetric case, large classes of two and three charge BPS solutions with the same asymptotic structure as five-dimensional black holes and black rings have been found. The solutions are typically constructed by first choosing a four-dimensional hyper-K\"ahler base space with non-trivial topology. One then constructs a five-dimensional supergravity solution by turning on magnetic fluxes on the non-trivial cycles of the base. These fluxes stabilize the two-cycles and are ultimately responsible for the non-trivial asymptotic charges. The homological two-cycles on the base ensure that there are no singular sources and the solutions can be made regular and causal. To argue in favor of the validity of Mathur's conjecture for non-supersymmetric, and non-extremal, black holes, one needs to construct a large number of similar smooth, horizonless gravity solutions which break supersymmetry and have the same charges and asymptotics as the black holes. There are very few solutions of this kind found so far, notable examples are the solutions of \cite{Jejjala:2005yu,Bena:2009qv}. Certainly it is of great interest to find more examples of such solutions and understand the possible implications for the resolution of black hole singularities and the information paradox.

Recently, there has been important progress in overcoming the difficulties of constructing exact non-BPS solutions of $\mathcal{N}=2$ five-dimensional supergravity \cite{Bena:2009qv,Goldstein:2008fq,Bena:2009ev,Bena:2009fi,Gimon:2007mh,Galli:2009bj,Ceresole:2009iy}. The underlying idea is to find a linear system of differential equations yielding non-supersymmetic solutions. Motivated by these advances the authors of \cite{Bena:2009fi} revisited the Ansatz and assumptions in the construction of BPS solutions to five-dimensional $\mathcal{N}=2$ supergravity coupled to vector multiplets \cite{Gauntlett:2002nw,Bena:2004de}. In this paper, the authors rederived the equations of motion, imposing a simple relation between the warp factor in the metric and the gauge fields, dubbed the ``floating brane" Ansatz. This Ansatz greatly simplifies the equations of motion and allows one not only to recover almost all known, BPS and non-BPS, classes of solutions, but also to find a new \textit{linear} system of equations. Using this result, new regular, horizonless and non-supersymmetric solutions were found in \cite{Bena:2009qv}. These solutions were constructed by solving the same linear system of equations as for BPS solutions, but on a Ricci-flat (instead of hyper-K\"ahler)  four-dimensional base. The particular examples discussed in \cite{Bena:2009qv} were based on the Euclidean Schwarzschild and Kerr-Taub-Bolt black holes. Our goal in this paper will be to construct new solutions based on the more general linear system of equations found in \cite{Bena:2009fi} and discuss their properties. Our solutions can be viewed as a generalization of the ones discussed in \cite{Bena:2009qv} since we consider four-dimensional base spaces which are electrovac solutions and are not not Ricci-flat.

We find a five-parameter family of smooth, horizonless solutions with a dyonic Euclidean Reissner-Nordstr\"om base. The solutions have general fluxes with no definite self-duality and are asymptotic to $\mathbb{R}^{1,3}\times S^1$. We generalize these solutions by including rotation and a NUT charge on the four-dimensional base, i.e. we use the Kerr-Newman-NUT background as a base. This more general family of solutions, still regular and horizonless, has six independent parameters, however their range is constrained by imposing regularity and causality of the five-dimensional background. Our solutions are not supersymmetric and have the same asymptotic structure as non-extremal black holes. They are therefore of interest, not only by themselves as new non-supersymmetric solutions, but also as candidates for microstates of non-extremal black holes. A general feature of the solutions is that the mass is linearly dependent on the electric charges. This property is due to the ``floating brane" Ansatz of \cite{Bena:2009fi}, which relates the warp factors in the five-dimensional metric to the electric gauge potentials. We also show that some of the solutions based on the Euclidean four-dimensional Kerr-Newman-NUT background exhibit ambipolar behavior: the four-dimensional base is allowed to have regions of positive and negative signature while the five-dimensional solution is everywhere completely regular and of definite Lorentzian signature. This provides some evidence that non-supersymmetric ambipolar solutions may also be ubiquitous like their BPS cousins  \cite{Bena:2005va,Berglund:2005vb}.

The solutions we find can be uplifted to solutions of eleven-dimensional supergravity on $T^6$. The construction is analogous to the the one used in \cite{Bena:2004de} to construct three-charge, five-dimensional BPS solutions. In this setup, our solutions have the same charge vector as three sets of M2 and M5 branes wrapping two- and four-cycles on the six-torus. The M2 branes will give the electric charges of the five-dimensional solution while the M5 branes, which also wrap a circle on the four-dimensional base, will be responsible for the dipole magnetic charges. It should be emphasized that our solutions will have no singular M2 and M5 brane sources. Because of the non-trivial topology of the four dimensional base the asymptotic charges of the solution are due to ``charges dissolved in fluxes''. This is essentially the same geometric transition mechanism as the one discussed in \cite{Bena:2005va} for BPS solutions. By using string dualities one can recast our solutions as six-dimensional solutions of IIB supergravity compactified on $T^4$ \cite{Bena:2008dw}. This duality frame may be useful for understanding the holographic dual field theory description of the solutions.

In Section 2 we present the action of five-dimensional $\mathcal{N}=2$ ungauged supergravity coupled to three $U(1)$ gauge fields and review the equations of motion and the Ansatz for our solutions. In Section 3 we find non-BPS supergravity solutions based on the four-dimensional Euclidean dyonic Reissner-Nordstr\"om black hole. In Section 4 we generalize this solutions to include rotation and a NUT charge in the four-dimensional base. Section 5 is devoted to conclusions and a discussion of possible extensions of our work. Finally, we discuss in Appendix A the extremal limits of the Reissner-Nordstr\"om and Kerr-Newman-Taub-Bolt backgrounds used in Sections 3 and 4 and their corresponding five-dimensional solutions.

\section{Equations of Motion}

\subsection{The five-dimensional Ansatz}

We will work with $\cN \! = \!  2$, five-dimensional ungauged supergravity with three $U(1)$ gauge fields and we use the conventions of \cite{Bena:2009fi}.  The bosonic action is
\begin{eqnarray}
  S = \frac {1}{ 2 \kappa_{5}} \int\!\sqrt{-g}\,d^5x \Big( R  -\coeff{1}{2} Q_{IJ} F_{\mu \nu}^I   F^{J \mu \nu} - Q_{IJ} \partial_\mu X^I  \partial^\mu X^J -\coeff {1}{24} C_{IJK} F^I_{ \mu \nu} F^J_{\rho\sigma} A^K_{\lambda} \bar\epsilon^{\mu\nu\rho\sigma\lambda}\Big) \,,
  \label{5daction}
\end{eqnarray}
with $I, J =1,2,3$. The scalars $X^I$ satisfy the constraint
\begin{equation} \label{Xcons}
X^1 X^2 X^3  = 1~,
\end{equation}
and there are therefore only two independent scalars. This is explained by the fact that one of the vector is in the gravity multiplet, and thus there are only two vector multiplets. For convenience, we introduce three other scalar fields, $Z_I$
\begin{equation}
X^1    =\bigg( \frac{Z_2 \, Z_3}{Z_1^2} \bigg)^{1/3} \,, \quad X^2    = \bigg( \frac{Z_1 \, Z_3}{Z_2^2} \bigg)^{1/3} \,,\quad X^3   =\bigg( \frac{Z_1 \, Z_2}{Z_3^2} \bigg)^{1/3}  \,.
\label{XZrelns}
\end{equation}
This automatically solves the constraint \eqref{Xcons}. The scalar kinetic term can be written as
\begin{equation}
  Q_{IJ} ~=~    \frac{1}{2} \,{\rm diag}\,\big((X^1)^{-2} , (X^2)^{-2},(X^3)^{-2} \big) \,.
\label{scalarkinterm}
\end{equation}
It is useful to introduce the scalar
\begin{equation}
Z ~\equiv~ \big( Z_1 \, Z_2 \, Z_3  \big)^{1/3}   \,.
\label{Zdefn}
\end{equation}
If one reduces the theory to four dimensions this will be a third independent scalar field. Having defined this new scalar, we will work with the following metric Ansatz
\begin{equation}
ds_5^2 ~=~ -Z^{-2} \,(dt + k)^2 ~+~ Z \, ds_4^2  \,,
\label{metAnsatz}
\end{equation}
We will denote the frames for (\ref{metAnsatz}) by $e^A$, $A=0, \dots,4$  and let $\hat e^a$, $a=1, \dots,4$ denote frames for $ds_4^2$.  Explicitly,
\begin{equation}
e^0 ~\equiv~     Z^{-1} \,(dt + k)\,, \qquad\qquad e^a ~\equiv~  Z^{1/2} \,\hat e^a \,.
\label{frames}
\end{equation}
We will assume also the ``floating brane'' Ansatz of \cite{Bena:2009fi}, which means that we take the metric coefficients to be related to the electrostatic potentials.  The Maxwell field is thus
\begin{equation}
A^{(I)}   ~=~  - Z_I^{-1}\, (dt +k) + B^{(I)}  \,,
\label{AAnsatz}
\end{equation}
where $B^{(I)}$ is a one-form on the base $ds_4^2$. Upon uplifting this solutions to eleven-dimensional supergravity, this Ansatz implies that M2 brane probes that have the same charges as the M2 branes sourcing the solution will have equal and opposite Wess-Zumino and Born-Infeld terms and hence will not feel any force.  Such brane probes may be placed anywhere in the base and may thus be viewed as ``floating.''

\subsection{Equations of motion}

The general equations of motion following from the above Ansatz were derived in \cite{Bena:2009fi} and we will use their results and conventions. We introduce the magnetic two-from field strengths
\begin{eqnarray}
	\Theta^{(I)} = d B^{(I)}~, 
\end{eqnarray}
and it will also be convenient to introduce the two-forms $\omega_-^{(I)}$ defined by
\begin{eqnarray}
	{1 \over 2}  \left( \Theta^{(I)} -  *_4 \Theta^{(I)}\right) \equiv C_{IJK} Z_J \omega_-^{(K)} \,,
\end{eqnarray}
where the $*_4$ is the Hodge dual with respect to the four-dimensional metric $ds_4^2$ in \eqref{metAnsatz}. Following \cite{Bena:2009fi} we will simplify the equations of motion by assuming
\begin{equation} \label{assumptions}
dk + *_4 dk = {1 \over 2} \sum_I Z_I\left( \Theta^{(I)} + *_4 \Theta^{(I)} \right)\,, \qquad \text{and} \qquad \omega^{(1)}_{-} = \omega^{(2)}_{-} = 0\,. 
\end{equation}
The four-dimensional base space has to be a solution of Euclidean Einstein-Maxwell theory\footnote{The normalization of the flux in this equation is different from most standard sources on general relativity and is chosen to agree with the four-dimensional conventions in \cite{Bena:2009fi}.} with (symbols with a $\hat{}$ live on the four-dimensional base)
\begin{equation} \label{electrovacequation}
\hat{R}_{\mu\nu} = \ds\frac{1}{2} \left( F_{\mu\rho}F_{\nu}^{\rho} - \ds\frac{1}{4} g_{\mu\nu} F_{\rho\sigma} F^{\rho\sigma} \right)\,,
\end{equation}
and
\begin{equation} \label{Fdecomp}
F = \Theta^{(3)} - \omega^{(3)}_{-}\,.
\end{equation}
The rest of the equations of motion reduce to\footnote{It is important to note that we have fixed the constant $\epsilon$ used in \cite{Bena:2009fi} to be $\epsilon=1$. This choice is not restrictive and it is straightforward to repeat all our calculations for $\epsilon=-1$.}
\begin{eqnarray}
	\hat{\nabla}^2 Z_1 = *_{4}(\Theta^{(2)}\wedge \Theta^{(3)})\,, \qquad\qquad (\Theta^{(2)} - *_{4} \Theta^{(2)}) = 2 Z_1 \, \omega_{-}^{(3)} \,, \label{EqZ1}
\end{eqnarray}
\begin{eqnarray}
	\hat{\nabla}^2 Z_2 = *_{4}(\Theta^{(1)}\wedge \Theta^{(3)})\,, \qquad\qquad (\Theta^{(1)} - *_{4} \Theta^{(1)}) = 2 Z_2 \, \omega_{-}^{(3)}\,, \label{EqZ2}
\end{eqnarray}
\begin{eqnarray}
\hat{\nabla}^2 Z_3 = *_{4}[\Theta^{(1)}\wedge \Theta^{(2)} - \omega_{-}^{(3)} \wedge (dk- *_{4}dk) ]\,, \label{EqZ3}
\end{eqnarray}
\begin{eqnarray}
dk+ *_{4}dk =  \ds\frac{1}{2}\ds\sum_{I=1}^{3} Z_{I}(\Theta^{(I)} + *_{4}\Theta^{I})\,. \label{Eqk}
\end{eqnarray}
An important point about this system of equations is that it can be solved in a linear fashion. In order to do that, one has to solve the equations in the right order. The starting point  is to choose a four-dimensional metric and its associated two-form field strength that solve \eqref{electrovacequation}. Then using \eqref{Fdecomp} one can read off $\Theta^{(3)}$ and $\omega_-^{(3)}$ from the field strength. Knowing these fields, \eqref{EqZ1} and \eqref{EqZ2} become systems of two linear coupled equations for $Z_1$ and $\Theta^{(2)}$ and $Z_2$ and $\Theta^{(1)}$ respectively. Finally, $k$ and $Z_3$ are solutions to the system of linear equations \eqref{EqZ3} and \eqref{Eqk}. We will show in the next sections how to solve these equations starting from the Euclidean Reisner-Nordstr\"om and Euclidean Kerr-Newman-NUT backgrounds. 

\section{Solutions with Euclidean Reissner-Nordstr\"{o}m base}

\subsection{The four-dimensional background}

Our starting point in this section will be the Euclidean dyonic Reissner-Nordstr\"{o}m background \cite{RN}
\begin{equation}
ds^2_{4} = \left(1-\ds\frac{2m}{r} + \ds\frac{p^2- q^2}{r^2}\right) d\tau^2 + \left(1-\ds\frac{2m}{r} + \ds\frac{p^2- q^2}{r^2}\right)^{-1} dr^2 + r^2 (d\theta^2+\sin^2\theta d\phi^2)~,
\end{equation}
\begin{equation}
F = \ds\frac{2q}{r^2} \, d\tau \wedge dr + 2 p \sin\theta \,d\theta \wedge d\phi \,.
\end{equation}
Where $m$ corresponds to the mass, $q$ to the electric charge and $p$ to the magnetic charge of the solution. This background solves the four-dimensional Einstein equations \eqref{electrovacequation}. It is useful to rewrite the metric as
\begin{equation}
ds^2_{4} =\ds\frac{(r-r_{+})(r-r_{-})}{r^2} d\tau^2 +\ds\frac{r^2}{(r-r_{+})(r-r_{-})} dr^2 + r^2 (d\theta^2+\sin^2\theta d\phi^2) \,.
\end{equation}
The constants $r_{\pm}$ are the Euclidean analogs of the inner and outer horizon of the Reissner-Nordstr\"{o}m black hole
\begin{equation}
r_{\pm} = m \pm \ds\sqrt{m^2 - p^2 + q^2} \,.
\end{equation}
To render $r_{\pm}$ real we restrict to the range of parameters\footnote{The case $m^2=p^2-q^2$ corresponds to the extremal Euclidean Reissner-Nordstr\"{o}m black hole. We discuss this case in Appendix A.} $m^2 > p^2-q^2$. Near the outer horizon one can set
\begin{equation}
r = r_{+} +  \ds\frac{r_{+} - r_{-}}{4r_{+}^2} \, \rho^2 \,, \qquad\qquad \chi = \ds\frac{r_{+} - r_{-} }{2r^2_{+}}\,\tau \,,
\end{equation}
and rewrite the metric as
\begin{equation}
ds^2_{NH} = d\rho^2 + \rho^2 d\chi^2 + r_{+}^2 (d\theta^2+\sin^2\theta d\phi^2) \,,
\end{equation}
which means that for a regular solution we should restrict to $r\geq r_{+}$ and the coordinate $\tau$ should be made periodic
\begin{equation}
\tau \sim \tau + \ds\frac{4\pi r_{+}^2}{r_{+}-r_{-}} \,.
\end{equation}
With this identification the metric is asymptotic to $\mathbb{R}^2\times S^2$ for $r \to r_{+}$ (i.e. we have a bolt of radius $r_{+}$ \cite{Gibbons:1979xm}) and to $\mathbb{R}^3\times S^1$ for $r\to \infty$. The angles $\theta$ and $\phi$ are the coordinates on $S^2$. In the next section we will solve the equations of motion of $\mathcal{N}=2$ five-dimensional supergravity with this Euclidean metric as a base space.

\subsection{The five-dimensional supergravity solution}

A convenient set of frames on the four-dimensional base is given by
\begin{eqnarray}
\hat{e}^1 &=&  \left(1-\ds\frac{2m}{r} + \ds\frac{p^2 - q^2}{r^2}\right)^{1/2}~ d\tau\,. \qquad\qquad \hat{e}^2 =  \left(1-\ds\frac{2m}{r} + \ds\frac{p^2 - q^2}{r^2}\right)^{-1/2}~ dr\,,\\
\hat{e}^3 &=&  r~ d\theta\,, \qquad\qquad \hat{e}^4 =  r\sin\theta~ d\phi\,,
\end{eqnarray}
and the usual self-dual and anti-self-dual two-forms are
\begin{equation}
\Omega_{\pm} = \hat{e}^{1} \wedge \hat{e}^{2} \pm \hat{e}^{3}\wedge \hat{e}^4\,.
\end{equation}
With this in hand it is easy to show that
\begin{equation}
\Theta^{(3)} = \ds\frac{p+q}{r^2}\,\Omega_{+}\,, \qquad\qquad \omega_{-}^{(3)} = \ds\frac{p-q}{r^2} \,\Omega_{-}\,.
\end{equation}
It will be useful to have the explicit expression for the potential $B^{(3)}$ satisfying $\Theta^{(3)}=d B^{(3)}$ 
\begin{eqnarray}\label{B3}
	B^{(3)} = {(p+q)\over r }\,d\tau -(p+q)\cos\theta \,d\phi \,.
\end{eqnarray}
The solution to equations \eqref{EqZ1} and \eqref{EqZ2} is
\begin{eqnarray}
Z_1 &=& 1 - \ds\frac{2q_2 (p+q)}{m}~ \ds\frac{1}{r}\,,\qquad\qquad Z_2 = 1 - \ds\frac{2q_1 (p+q)}{m}~ \ds\frac{1}{r}\,,\\
\Theta^{(1)} &=& f_1(r) \Omega_{+} + g_1(r) \Omega_{-}\,, \qquad\qquad \Theta^{(2)} = f_2(r) \Omega_{+} + g_2(r) \Omega_{-}\,,
\end{eqnarray}
where
\begin{eqnarray}
f_1 &=& \ds\frac{2q_1}{r^2} - \ds\frac{2q_1(p^2-q^2)}{m \,r^3}\,, \qquad\qquad f_2 = \ds\frac{2q_2}{r^2} - \ds\frac{2q_2(p^2-q^2)}{m\, r^3}\,,\\
g_1 &=& \ds\frac{(p-q)}{r^2} - \ds\frac{2q_1(p^2-q^2)}{m \, r^3}\,,\qquad\qquad g_2 = \ds\frac{(p-q)}{r^2} - \ds\frac{2q_2(p^2-q^2)}{m \, r^3}\,.
\end{eqnarray}
Note that with these functions $f_I(r)$ and $g_I(r)$ one can show that $d\Theta^{(I)}=0$, which means that locally one can express $\Theta^{(1)}$ and $\Theta^{(2)}$ in terms of potential one-forms, $\Theta^{(I)}=dB^{(I)}$. Explicitly, these one-forms are
\begin{eqnarray}
	B^{(I)}=K_I \,d \tau  + b_I \,d \phi \, , 
\end{eqnarray}
with
\begin{eqnarray} \label{B1}
	K_1 &=& {2q_1 + p-q \over r} - {2q_1(p^2-q^2) \over m \, r^2} \,, \qquad b_1 = (-2 q_1 + p-q)\cos \theta \,, \\ \label{B2} 
	K_2 &=& {2q_2 + p-q \over r} - {2q_2(p^2-q^2) \over m \, r^2} \,, \qquad b_2 = (-2 q_2 + p-q)\cos \theta \,. 
\end{eqnarray}
To solve \eqref{EqZ3} and \eqref{Eqk}, we will use the Ansatz
\begin{equation}
k = \mu(r) d\tau + \nu(\theta) d\phi\,.
\end{equation}
One can then show that
\begin{equation}
\nu(\theta) = \nu_{0} + \xi \cos\theta\,,
\end{equation}
with $\nu_0$ and $\xi$ constants. Then the problem reduces to a system of two coupled \textit{linear} ordinary differential equatons for $\mu(r)$ and $Z_3(r)$
\begin{eqnarray}
\ds\frac{d\mu}{dr} &=& - \left( \ds\frac{\xi}{r^2} +Z_1f_1 + Z_2f_2 + \ds\frac{p+q}{r^2} Z_3  \right) \,, \\
\hat{\nabla}^2 Z_3 &=& 2 \left( f_1f_2 - g_1g_2 + \ds\frac{\xi (p-q)}{r^4} - \ds\frac{(p-q)}{r^2}~\ds\frac{d\mu}{dr}  \right) \,.
\end{eqnarray}
A solution to these equations is given by
\begin{eqnarray}
Z_3 &=& 1 - \left( \ds\frac{4q_1q_2 (m^2-p^2+q^2)}{m^3} + \ds\frac{2(p-q)(q+q_1+q_2)}{m} \right)~ \ds\frac{1}{r} + \ds\frac{4q_1q_2 (p^2-q^2)}{m^2}~\ds\frac{1}{r^2}\,,\\
\mu &=& (p+q+2(q_1+q_2)) \left( \ds\frac{1}{r} - \ds\frac{1}{r_{+}} \right) \notag\\&& - \left( \ds\frac{2q_1q_2(p+q)(3m^2-p^2+q^2)}{m^3} +\ds\frac{(p^2-q^2) (q+2q_1+2q_2)}{m} \right) \left( \ds\frac{1}{r^2} - \ds\frac{1}{r_{+}^2} \right) \notag\\&&+ \ds\frac{4 q_1q_2 (p^2-q^2) (p+q)}{m^2} \left( \ds\frac{1}{r^3} - \ds\frac{1}{r_{+}^3} \right)\,.
\end{eqnarray}
To arrive at this particular solution we have chosen 
\begin{equation}
\nu_0 = \xi=0\,, \qquad \to \qquad \nu = 0\,,
\end{equation}
which ensures that there are no closed time-like curves (CTCs) coming from the $d\phi^2$ term in the five-dimensional metric, at $\theta=0,\pi$. We have also chosen the additive constant in the solution for $\mu$ such that $\mu(r_{+})=0$, which ensures the absence of CTCs  near the bolt. This implies that $\mu$ has a non vanishing value $\gamma$ at infinity, 
\begin{eqnarray}
	\lim_{r\to \infty}\mu = \gamma \equiv  &&- \ds\frac{1}{r_{+}}(p+q+2(q_1+q_2)) \\ \nonumber 
	 &&+  \ds\frac{1}{r_{+}^2}  \left( \ds\frac{2q_1q_2(p+q)(3m^2-p^2+q^2)}{m^3} +\ds\frac{(p^2-q^2) (q+2q_1+2q_2)}{m} \right)  \\ \nonumber
	 &&- \ds\frac{1}{r_{+}^3} \ds\frac{4 q_1q_2(p^2-q^2)(p+q)}{m^2}  \,,
\end{eqnarray}
this will be important in the calculation of the asymptotic charges of the five-dimensional solution. Note also that we have set the constants terms in $Z_I$ to $1$ by which we fix the asymptotic values of the scalar fields\footnote{In an eleven-dimensional uplift of our solution this choice will fix the asymptotic volumes of the two-cycles of $T^6$.}.

An important difference between this solution and the magnetized Euclidean Schwarzschild solution in \cite{Bena:2009qv} is that the fluxes here are not self-dual. It is clear that if we set 
\begin{equation}
q=p= \ds\frac{\tilde{q}_3}{2}\,, \qquad\qquad q_1 = \ds\frac{\tilde{q}_1}{2}\,, \qquad\qquad q_2=\ds\frac{\tilde{q}_2}{2}\,,
\end{equation}
we will recover the five-dimensional solution based on the Euclidean Schwarzschild black hole found in \cite{Bena:2009qv}. Note that all $q_I$ in \cite{Bena:2009qv} should be identified with $\tilde{q}_I$, this is due to the different conventions in the normalization of the fluxes.

\begin{figure}[t]
 \centering
    \includegraphics[width=10cm]{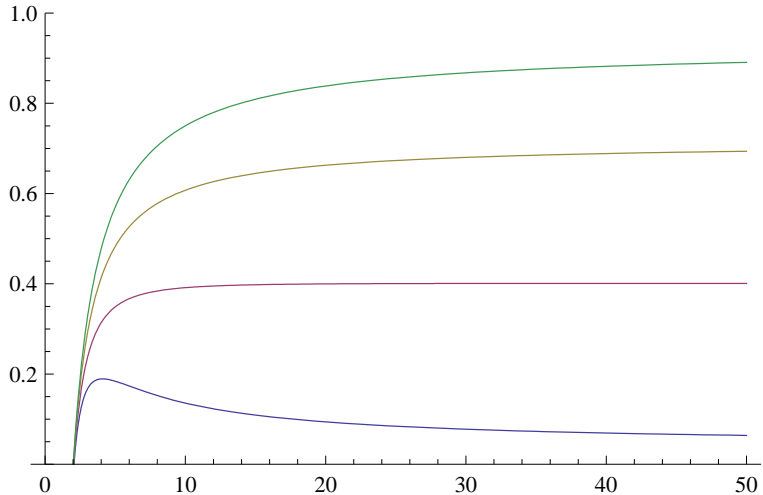}
    \caption{\it \small $\mathcal{M}$ as a function of $\rho=r/r_{+}$ for four different values of $Q/m$. The curves correspond to $Q/m=(0.1,0.2,0.3,0.4)$ from top to bottom. }
\label{fig1}
\end{figure}

An important step in the analysis of the five-dimensional solution constructed above is to ensure the global absence of CTCs. This means that for constant time slices one should make sure that the coefficient of $d\tau^2$ in the five-dimensional metric is non-negative and all $Z_{I}$ are positive definite. To analyze this condition in an explicit example we will take
\begin{equation}
q=q_1=q_2 = Q>0\,, \qquad\qquad p = \ds\frac{Q}{2}\,. \label{parametersRN}
\end{equation}
Then we have 
\begin{equation}
r_{\pm} = m \pm \sqrt{m^2+\ds\frac{3Q^2}{4}}\,,
\end{equation}
and the condition that $Z_1$ and $Z_2$ are positive for $r \geq r_{+}$ imposes
\begin{equation}
0< \ds\frac{Q}{m} < \ds\frac{\sqrt{3}}{2}  \approx 0.8660\,.
\end{equation}
Requiring that $Z_3$ is positive for $r>r_{+}$ leads to 
\begin{equation}
0< \ds\frac{Q}{m} \lessapprox 0.7783\,,
\end{equation}
which is clearly a stronger constraint.
Finally we have to make sure that the coefficient of $d\tau^2$ is non-negative
\begin{equation}
\mathcal{M} \equiv \ds\frac{1}{r^2 (Z_1Z_2Z_3)^{2/3}} [  Z_1Z_2Z_3 (r-r_{+}) (r-r_{-}) - \mu^2r^2 ] \geq0\,.
\end{equation}
Expanding this expression for $r \to \infty$ we find a sextic algebraic inequality in $Q/m$, which can be solved numerically. The allowed range of parameters coming from this constraint is
\begin{equation}
0 < \ds\frac{Q}{m} \lessapprox 0.4118 \,, \qquad\qquad 0.8811 \lessapprox ~\ds\frac{Q}{m}~ \lessapprox 1.2587\,.
\end{equation}
The bottom line is that for the choice of parameters \eqref{parametersRN} the five-dimensional solution is completely regular and there are no CTCs (globally) if
\begin{equation}
0 < \ds\frac{Q}{m} \lessapprox 0.4118 \,.
\end{equation}
Some plots of $\mathcal{M}$ for different values of $Q/m$ are presented in Figure 1. We have performed a detailed numerical analysis for a number of other choices for the parameters $(p,q,q_1,q_2)$ and the conclusions are qualitatively the same. Namely, there is a region in parameter space in which the five-dimensional solution is regular and has no global CTCs.

\subsection{The asymptotic charges}
\label{sec:chargeRN}

Having found a regular five-dimensional solution of $\mathcal{N}=2$ ungauged supergravity, asymptotic to $\mathbb{R}^{1,3}\times S^1$, it is instructive to compute its asymptotic charges. The dipole charges , $d_I$, of the solution are directly encoded in the magnetic part of the gauge field, $B^{(I)}$. We thus have from (\ref{B3}), (\ref{B1}) and (\ref{B2}) 
\begin{eqnarray}
	d_1 &=& 2q_1-p+q \,, \nonumber \\
	d_2 &=& 2q_2-p+q \,, \\ \nonumber
	d_3 &=& p+q \,.  
\end{eqnarray}
If the solution is viewed as a compactification of eleven-dimensional supergravity on $T^6$ these will correspond to the M5 brane charges. The electric charges of the solution are given by
\be
Q_I =  \int_{S^1\times S^2}  \Bigl[(X^I)^{-2} *_5 d A^I-\coeff{1}{2} C_{IJK} A^J \wedge d A^K\Bigr]\,,
\label{QI}
\ee
where the integral is computed over the $S^1\times S^2$ at spatial infinity, parameterized by $(\tau,\theta,\phi)$. The Chern-Simons term gives a non-vanishing contribution to the charge, due to the fact that the one-form $k$ goes to a constant non-zero value at infinity. A straightforward calculation yields
\bea
 Q_1 \!\!\!&=&\!\!\! - {16\pi^2 r_+^2 \over r_+ -r_-} \left(  {2(p+q)q_2 \over  m} + \gamma (q+q_2)\right)\,, \notag\\
 Q_2 \!\!\!&=&\!\!\! - {16\pi^2 r_+^2 \over r_+ -r_-} \left(  {2(p+q)q_1 \over  m} + \gamma (q+q_1)\right)\,, \\
 Q_3 \!\!\!&=&\!\!\! - {16\pi^2 r_+^2 \over r_+ -r_-} \left(  {4 q_1 q_2 \over  m} + \gamma (q_1+q_2+p-q) + {2(p-q)(q+q_1+q_2) \over m} - {4q_1q_2(p^2-q^2) \over m^3} \right)\,. \notag
\label{charges}
\eea
To compute the mass and the Kaluza-Klein (KK) electric charge of the solution one has
to analyze the asymptotic form of the metric. The fact that the
one-form, $k$, does not vanish at infinity implies that the
coordinates $(\tau,t)$ define a frame which is not asymptotically at
rest. One can go to an asymptotically static frame by
casting the large $r$ limit of the metric in the form
\be
ds^2\approx  (1-\gamma^2)\Bigl(d\tau - {\gamma\over 1-\gamma^2} dt \Bigr)^2 - {1 \over 1-\gamma^2} dt^2 + dr^2 + r^2 (d\theta^2+\sin^2\theta d\phi^2)\,,
\ee
and redefining the coordinates as
\be
\hat{\tau}=(1-\gamma^2)^{1/2} \Bigl(\tau -{\gamma\over 1-\gamma^2} t\Bigr)\,,\qquad\qquad \hat{t}=(1-\gamma^2)^{-1/2} t\,.
\label{hatcoord}
\ee
To compute the mass and KK charge, one needs to reduce our solution along the the $\hat \tau $ coordinate. The metric takes the form 
\bea
ds^2_5={g^2 \over  Z^{2}} \hat{I}_4\Bigl[d\hat \tau + \left(\gamma - {\mu \over g^2 \hat{I}_4}\right) d\hat t \Bigr]^2+ {Z \over g \hat{I}_4^{1/2}} ds^2_E\,,
\eea
where we have defined, 
\be
g=1-{2m\over r}+{p^2-q^2 \over r^2}~, \qquad\qquad \hat{I}_4= {1 \over 1-\gamma^2} \left( g^{-1}Z^3 -g^{-2}\mu^2 \right) \,,
\ee
and
\be
ds^2_E= -\hat{I}_4^{-1/2} d\hat{t}^2 + \hat{I}_4^{1/2} \Bigr[dr^2 + g r^2 (d\theta^2 + \sin^2\theta d\phi^2)\Bigr]
\ee
is the four-dimensional Einstein metric. From the asymptotic behavior of the $d\hat{t}^2$ coefficient in the Einstein frame metric one can read off the mass
of the solution
\bea
 M &=& {1\over G_4(1- \gamma^2)}\Bigl[ {m \over 2}(1-2\gamma^2) - {q_1 q_2 + p q_1 + p q_2 + {q (p-q) \over 2} \over m} \\ \nonumber
&& \qquad - \gamma (q_1+q_2+ {p+q \over 2} ) + {q_1 q_2 (p^2-q^2) \over m^3} \Bigr]\,.
\eea
Here $G_4$ is the four-dimensional Newton's constant, whose relation to the five-dimensional Newton's constant $G_5$ is
\bea
G_4 = \ds\frac{G_5}{\text{vol}(\tau)} = {G_5 \over (1-\gamma^2)^{1/2}} {(r_+ -r_{-}) \over 4\pi r_+^2}\,,
\eea
and $\text{vol}(\tau)$ is the length of the $S^1$ parametrized by $\tau$. The KK electric charge, $Q_e$, is encoded in the KK gauge field\footnote{We use the conventions of \cite{Elvang:2005sa}.}
\be
A_{KK}=  \left(\gamma - { \mu \over g^2 \hat{I}_4 }\right) d\hat{t}~,
\ee
and is given by
\bea
 Q_e&=&-{1\over G_4(1-\gamma^2)}\Bigl[ \gamma {m \over 2} + \gamma  {q_1 q_2 + p q_1 + p q_2 + {q (p-q) \over 2}\over m}  + {1+\gamma^2 \over 2}(q_1+q_2+ { p+q \over 2}) \\ \nonumber
&& \qquad   -~ \gamma{q_1 q_2 (p^2-q^2) \over m^3} \Bigr]\,.
\eea
Finally it is instructive to compute the rest-mass, $M_0$, of the solution, i.e. the mass with respect to the ($t,\tau$) frame
\be
	M_0 \equiv (1-\gamma^2)^{-1/2}(M-\gamma Q_e) = {1 \over 16 \pi G_5}\left({32 \pi^2 r_+^2 m \over r_+ - r_-} +Q_1+Q_2+Q_3\right)\,.
\label{mass}
\ee

It is clear from this expression that if we set the mass of the four-dimensional Reissner-Nordstr\"om black hole to zero we will recover the usual relation between the mass and the charges of a BPS black hole solution. Note also that despite the fact that we start our construction from a four-dimensionnal black hole with a magnetic charge $p$, $A_{KK}$ has a component only along $d \hat t$, which implies that the final solution does not carry any global magnetic charge.

\section{Adding rotation and NUT charge}

\subsection{The four-dimensional background}

We now generalize the four-dimensional Euclidean base from the previous section to include an angular momentum parameter $\alpha$ and a NUT charge $N$. The metric and the two-form flux are
\begin{eqnarray} \label{KNTBmetric} 
ds^2_{4} &=& \ds\frac{\Sigma}{\Delta} \, dr^2 + \Sigma \, d\theta^2 + \ds\frac{\sin^2\theta}{\Sigma} \, (\alpha d\tau+P_r d\phi)^2 + \ds\frac{\Delta}{\Sigma} \, (d\tau+P_{\theta}d\phi)^2\,,  \label{4DKNTBmetric} \\
F &=& \ds\frac{p+q}{[r-(N+\alpha\cos\theta)]^2} \, \Omega_{+} - \ds\frac{p-q}{[r+(N+\alpha\cos\theta)]^2} \, \Omega_{-} \,, \label{4DKNTB2form}
\end{eqnarray}
where we defined the functions
\begin{eqnarray}
P_r &=& r^2 - \alpha^2 - \ds\frac{N^4}{N^2-\alpha^2}\,, \qquad\qquad P_{\theta} = 2N\cos\theta - \alpha\sin^2\theta - \ds\frac{\alpha N^2}{N^2-\alpha^2} \,, \\
\Delta &=& r^2 - 2mr+ N^2 - \alpha^2   + p^2 - q^2\, \qquad  \qquad \Sigma = P_r - \alpha P_{\theta} = r^2 - (N + \alpha\cos\theta)^2 \,.
\end{eqnarray}
The (anti-)self-dual two-forms $\Omega_{\pm}$ are 
\begin{equation}
\Omega_{\pm} = \hat{e}^1\wedge \hat{e}^4 \pm \hat{e}^2\wedge \hat{e}^3\,,
\end{equation}
with the four-dimensional vielbeins
\begin{eqnarray}
\hat{e}^1 &=& \left(\ds\frac{\Sigma}{\Delta}\right)^{1/2} dr\,, \qquad\qquad\qquad \hat{e}^2 = \left(\Sigma\right)^{1/2} d\theta\,,\\
\hat{e}^3 &=& \ds\frac{\sin\theta}{\left(\Sigma\right)^{1/2}} (\alpha d\tau + P_{r}d\phi)\,, \qquad\qquad \hat{e}^4 =  \left(\ds\frac{\Delta}{\Sigma}\right)^{1/2}  ( d\tau + P_{\theta}d\phi)\,.
\end{eqnarray}
The four-dimensional metric (\ref{4DKNTBmetric}) and gauge field (\ref{4DKNTB2form}) are solutions to the Einstein-Maxwell equations (\ref{electrovacequation}). The parameters $m$, $q$ and $p$ still correspond respectively to the mass, electric charge and magnetic charge of the four-dimensional Euclidean solution. The new parameters are the NUT charge, $N$, and the angular momentum parameter, $\alpha$. This background is a generalization of the familiar Kerr-Newman solution \cite{Kerr:1963ud} to which we have added magnetic and NUT charges. Note also that the Kerr-Newman-NUT metric \eqref{KNTBmetric} has exactly the same form as the Kerr-Taub-Bolt metric \cite{Gibbons:1979nf}, the only difference is in the function $\Delta$. One can recover the Kerr-Taub-Bolt metric of \cite{Gibbons:1979nf} by taking $p=q$. The Euclidean analogs of the inner and outer horizon of the black hole are given by the zeroes of $\Delta$   
\begin{eqnarray}
	\Delta = (r- r_+)(r-r_-) \,, \qquad\qquad r_{\pm} =  m \pm \sqrt{m^2 - N^2 + \alpha^2 - p^2 + q^2}\,.
\end{eqnarray}
The analysis of the regularity of this four-dimensional background is exactly the same as the one performed in \cite{Bena:2009qv}, \cite{Gibbons:1979nf} for the Kerr-Taub-Bolt solution. We will not reproduce it here and will present only the conclusions. We are interested in the case where the roots $r_{\pm}$ of $\Delta$ are real, in order to have a non-trivial bolt. This imposes
\be \label{rplusreal}
	m^2  \geq N^2 - \alpha^2 + (p^2-q^2)\,.
\ee
Then, the metric is regular provided that 
\begin{eqnarray}
	r \geq r_+ \,, \qquad \phi \sim \phi + 2\pi \,, \qquad \tau \sim \tau + 8 \pi N \sim \tau + {2\pi \over \kappa} \,, 
\end{eqnarray}
where we defined
\begin{eqnarray}
	P_{r+} \equiv P_r(r=r_+) = r_+^2 - \alpha^2 - {N^4 \over N^2 - \alpha^2} \,, \qquad \kappa \, \equiv \,  \, \Big|{r_+ - r_- \over 2 P_{r+}}\Big|  \,. 
\end{eqnarray}
\begin{figure}[t]
 \centering
    \includegraphics[width=8cm]{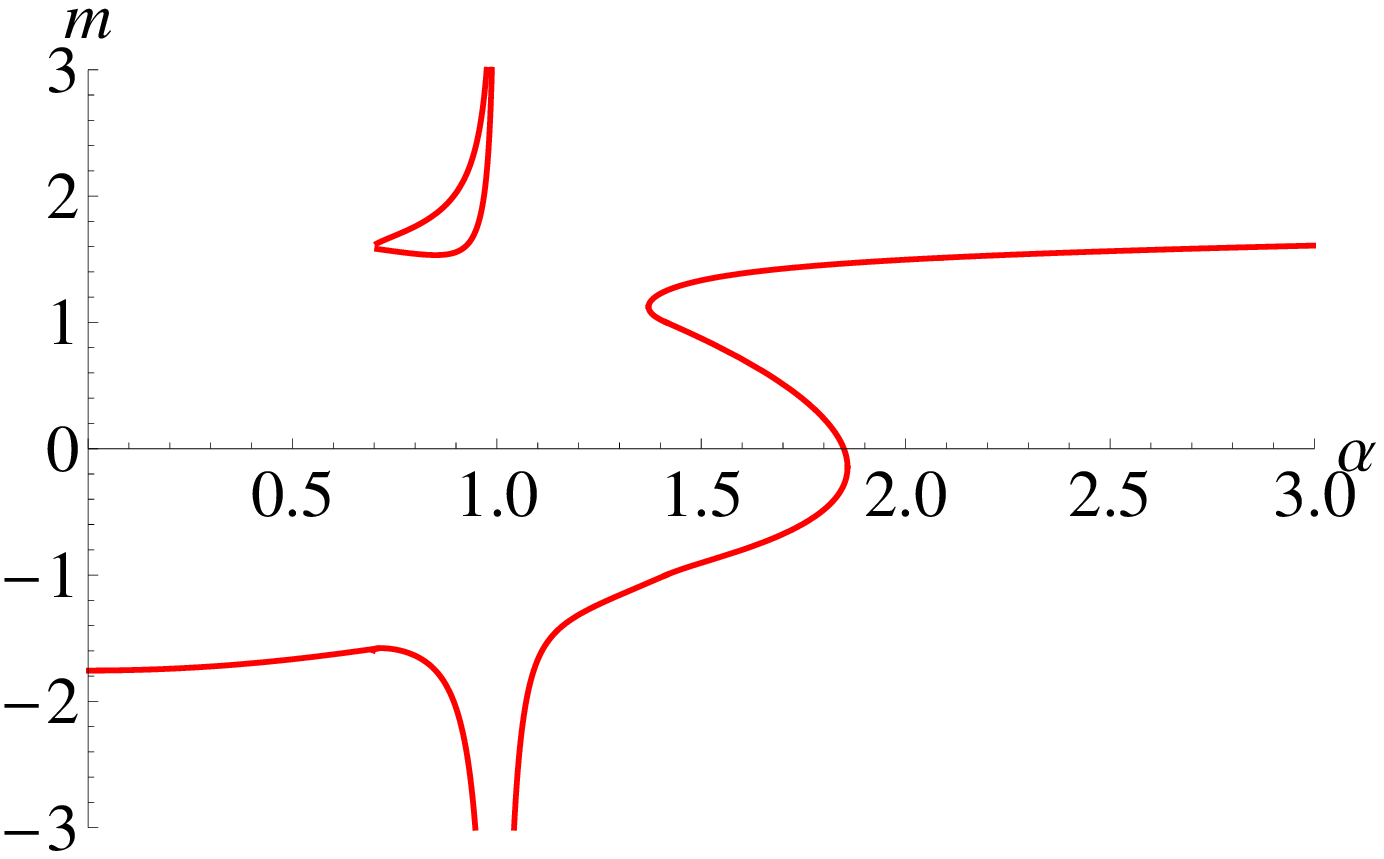}
  \hfill
    \includegraphics[width=8cm]{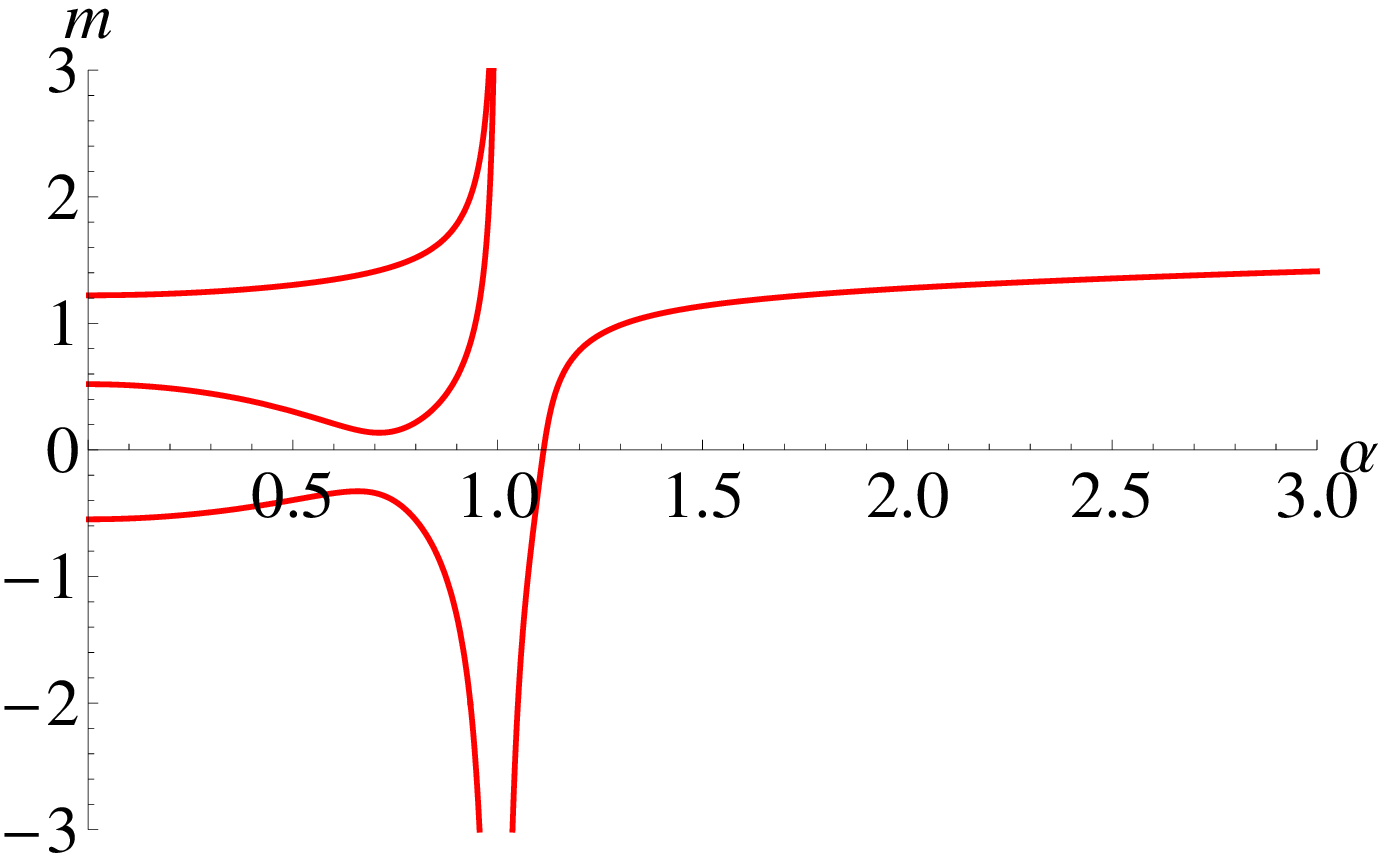}
    \caption{{\it \small The two graphs represented here are plots of $m$ as a function of $\alpha$, in units in which $N=1$ (this choice can always be made because the equations are homogeneous). They show the solutions to \eqref{paramrelation} for $p^2-q^2=2$ (left) and $p^2-q^2=-3/4$ (right). As the value of $p^2-q^2$ changes, the different branches of the solution evolve and some non trivial differences can be seen. For example, for $p^2-q^2=2$, one can see that there is only one possible value of $m$ for $\alpha=0$, in contrast with the three different possibilities for $p^2-q^2=-3/4$. The important feature is that for any given value of $p^2-q^2$, there will always be a solution to \eqref{paramrelation}.}}  
\label{fig2}
\end{figure}
Regularity imposes two \textsl{a priori} independent periodicities for the coordinate $\tau$: $\tau \sim \tau + 8\pi N $ comes from imposing regularity for $r \to \infty$ and $\tau \sim \tau + {2\pi \over \kappa} $ is a regularity condition at $r=r_{+}$. To have a globally regular four-dimensional base with no conical singularities we have to impose the following constraint
\begin{eqnarray}\label{paramrelation}
	\kappa = {1 \over 4 |N|} \,.
\end{eqnarray}
It is imporant to mention that if we want this metric to have signature ($+$,$+$,$+$,$+$), in order for it to be a regular Euclidean four-dimensional metric, one has to impose that $\Sigma$ remains positive, and this will restrict the allowed range of parameters. However, since we are interested here in constructing a regular five-dimensional solution starting from a four-dimensional base, we do not have to impose that the four-dimensional signature stays positive. The only requirement is that we end up with a regular Lorentzian five-dimensional solution,  we will discuss this point in section \ref{ambipolar}. Therefore, the physical constraints on the parameters of the solutions are \eqref{rplusreal} and \eqref{paramrelation}. Before constructing the five-dimensional solution, it is worth analyzing what \eqref{paramrelation} imposes on the parameters $m$, $N$, $\alpha$, $p$ and $q$. 

One can easily see, using the definition of $\kappa$, that \eqref{paramrelation} only involves $|N|$ and $|\alpha|$. We will therefore assume $N$ and $\alpha$ to be positive to study this constraint. Note also that $p$ and $q$ only appear in the combination $p^2-q^2$. In order to solve \eqref{paramrelation}, the simplest approach is to get rid of the square roots in \eqref{paramrelation}, and this gives a constraint that is cubic in $m$, and quadratic in $p^2-q^2$. This constraint depends on the sign of $P_{r+}$: if $P_{r+}$ is positive, we have
\begin{eqnarray}\label{cubic1}
&& 16 N(N^2 - \alpha^2)^2 m^3 - 4(N^2-\alpha^2)(5N^4 -3 N^2\alpha^2 - \alpha^4) m^2  \notag \\ 
&& - 16 N(N^2 - \alpha^2)^2(N^2 - \alpha^2 + p^2 -q^2) m + 20 N^8 - 52 N^6\alpha^2 + 49 N^4\alpha^4 - 16N^2\alpha^6  \notag \\
&& + 2N^2(p^2-q^2)(N^2-\alpha^2)(10N^2-9\alpha^2) + (p^2-q^2)^2(N^2-\alpha^2)^2~=~ 0 \,,
\end{eqnarray}
if $P_{r+}$ is negative equation (\ref{paramrelation}) implies
\begin{eqnarray}\label{cubic2}
&& -16 N(N^2 - \alpha^2)^2 m^3 - 4(N^2-\alpha^2)(5N^4 -3 N^2\alpha^2 - \alpha^4) m^2  \notag \\
&& + 16 N(N^2 - \alpha^2)^2(N^2 - \alpha^2 + p^2 -q^2) m + 20 N^8 - 52 N^6\alpha^2 + 49 N^4\alpha^4 - 16N^2\alpha^6  \notag\\ 
&& + 2N^2(p^2-q^2)(N^2-\alpha^2)(10N^2-9\alpha^2) + (p^2-q^2)^2(N^2-\alpha^2)^2~=~ 0 \,,
\end{eqnarray}
which is the same as (\ref{cubic1}) but with $ m \to -m$.  Note that a solution to \eqref{cubic1} or \eqref{cubic2} is not automatically a solution to (\ref{paramrelation}). Indeed, one has first to make sure to solve either \eqref{cubic1} or \eqref{cubic2} in the domains where $P_{r+}$ is respectively positive or negative; secondly, by squaring the square roots, one has to insure that the expression to which this square root is equal is positive. We performed a detailed analysis of these relations for many different values of the parameters, including $p^2-q^2$. Our analysis shows that, even if the explicit form of the branches of the solutions can differ quite a lot, there are solutions to \eqref{paramrelation} for any value of $p^2-q^2$. For illustration, we present in Figure \ref{fig2} the solution to \eqref{paramrelation} for two different values of $p^2-q^2$.

\subsection{The five-dimensional supergravity solution}

We can use the regular four-dimensional electrovac solution from the previous section to construct a five-dimensional supergravity solution by solving the equations from Section 2.2. From the four-dimensional solution one can read off
\begin{equation}
\Theta^{(3)} = \ds\frac{p+q}{[r-(N+\alpha\cos\theta)]^2} \Omega_{+} \, , \qquad\qquad \omega_{-}^{(3)} = \ds\frac{p-q}{[r+(N+\alpha\cos\theta)]^2} \Omega_{-}\,.
\end{equation}
These two-forms are $d$-closed, and thus (at least locally) have corresponding one-form potentials
\begin{eqnarray}
	\Theta^{(3)} = (p+q) \, d A_+ \,, \qquad\qquad \omega_-^{(3)} = (p-q) \, d A_- \,,
\end{eqnarray}
which are given by
\begin{eqnarray}
	A_{\pm} = -{1 \over r \mp(N+\alpha \cos\theta) } \, (d\tau + P_{\theta} d\phi) \mp \cos \theta \, d\phi \,.
\end{eqnarray}
We now want to solve \eqref{EqZ1} and \eqref{EqZ2}. As noted above, once we know the four-dimensional base space, $\Theta^{(3)}$ and $\omega_-^{(3)}$, \eqref{EqZ1} is a coupled system of two linear equations for $Z_1$ and $\Theta^{(2)}$. Defining
\begin{eqnarray}
	\Theta^{(2)} = f_2(r,\theta)\,\Omega_+ + g_2(r,\theta)\,\Omega_- \,,
\end{eqnarray}
\eqref{EqZ1} can be rewritten as
\begin{eqnarray}
	\hat \nabla^2 Z_1 \!\!&=&\!\! {2 f_2 (p+q) \over [r-(N+\alpha\cos\theta)]^2} \,, \\
	g_2 \!\!&=&\!\!\ds\frac{p-q}{[r+(N+\alpha\cos\theta)]^2} Z_1\,. \nonumber 
\end{eqnarray}
The solution to this system is given by
\begin{eqnarray}
Z_1 &=& 1-  \ds\frac{2q_2(p+q)}{m-N}\ds\frac{1}{r-(N+\alpha\cos\theta)}\,, \\
f_2 &=& \ds\frac{2q_2}{[r-(N+\alpha\cos\theta)]^2} -  \ds\frac{2q_2(p^2-q^2)}{m-N}\ds\frac{1}{[r-(N+\alpha\cos\theta)]^2[r+(N+\alpha\cos\theta)]}\,,\\
g_2 &=& \ds\frac{p-q }{[r+(N+\alpha\cos\theta)]^2} - {2q_2(p^2-q^2) \over m-N}\ds\frac{1}{[r-(N+\alpha\cos\theta)][r+(N+\alpha\cos\theta)]^2} \,.
\end{eqnarray}
Similarly, \eqref{EqZ2} is solved by
\begin{eqnarray}
Z_2 &=& 1-  \ds\frac{2q_1(p+q)}{m-N}\ds\frac{1}{r-(N+\alpha\cos\theta)}\,,\\
f_1 &=& \ds\frac{2q_1}{[r-(N+\alpha\cos\theta)]^2} -  \ds\frac{2q_1(p^2-q^2)}{m-N}\ds\frac{1}{[r-(N+\alpha\cos\theta)]^2[r+(N+\alpha\cos\theta)]}\,,\\
g_1 &=&\ds\frac{p-q}{[r+(N+\alpha\cos\theta)]^2} -{2q_1(p^2-q^2) \over m-N}\ds\frac{1}{[r-(N+\alpha\cos\theta)][r+(N+\alpha\cos\theta)]^2} \,,
\end{eqnarray}
and $q_1$ and $q_2$ are constants related to the electric charges of the solution\footnote{Note that, as in the Reissner-Nordstr\"om solution, our $q_I$ differ from the ones in \cite{Bena:2009qv} by a factor of 2.}. 

One can show that the 2-forms $\Theta^{(I)}$, $I=1,2$, are $d$-closed, and the corresponding one form potentials, $B^{(I)}$, are given by 
\begin{eqnarray}
	B^{(I)} = 2q_I \, A_+ + (p-q) \, A_- + {2q_I (p^2-q^2) \over m-N}{1\over \Sigma} \, (d\tau + P_{\theta}\, d\phi) \,.
\end{eqnarray}
We now have to solve the last system of equations \eqref{EqZ3}, \eqref{Eqk}, to find $Z_3$ and the angular momentum, $k$, of the solution. We choose the following Ansatz for $k$
\begin{equation}
k = \mu(r,\theta) \, (d\tau + P_{\theta} d\phi) + \nu(r,\theta) \, d\phi\,.
\end{equation}
After some work one finds
\begin{multline}
dk = \left( \partial_{r}\mu -\ds\frac{\alpha}{\Sigma} \partial_{r}\nu \right)\hat{e}^1\wedge \hat{e}^4 + \ds\frac{\Delta^{1/2}}{\Sigma\sin\theta} \partial_{r}\nu ~\hat{e}^1\wedge \hat{e}^3 \\+ \ds\frac{1}{\Sigma\sin\theta} (\mu \, \partial_{\theta}P_{\theta} +\partial_{\theta}\nu) ~\hat{e}^2\wedge \hat{e}^3 + \ds\frac{\partial_{\theta}(\Sigma\mu-\alpha\nu)}{\Sigma\,\Delta^{1/2}}~ \hat{e}^2\wedge \hat{e}^4~.
\end{multline}
Equation \eqref{Eqk} imposes a relation between the functions $\mu$ and $\nu$
\begin{eqnarray}
	\Delta \partial_r \nu = \sin\theta \, \partial_{\theta}(\Sigma \mu - \alpha \nu ) \,.
\end{eqnarray}
Using this constraint one can express $\mu$ and $\nu$ in terms of a single function $F(r,\theta)$ as
\begin{equation}
\mu = \ds\frac{\Delta~\partial_{r}F + \alpha\sin\theta \,\partial_{\theta}F}{\Sigma}\,, \qquad\qquad \nu = \sin\theta \, \partial_{\theta}F\,. \label{muandnuKN}
\end{equation}
With this in mind one can rewrite \eqref{EqZ3} and \eqref{Eqk} as
\begin{eqnarray}
&&\mathcal{D}_{+}F =  Z_1f_1 + Z_2f_2 + \ds\frac{(q+p)Z_3}{[r-(N+\alpha\cos\theta)]^2}\,, \label{Eqk2} \\
&&\hat{\nabla}^2 Z_3 = 2(f_1f_2-g_1g_2) + \ds\frac{2(p-q)}{[r+(N+\alpha\cos\theta)]^2}\mathcal{D}_{-}F \,, \label{EqZ32}
\end{eqnarray}
where we have defined
\begin{eqnarray}
\mathcal{D}_{\pm}F = \ds\frac{1}{\Sigma} \left[ \partial_{r}(\Delta \partial_{r} F) \pm \ds\frac{\partial_{\theta}(\sin\theta \, \partial_{\theta}F)}{\sin\theta} - \ds\frac{2}{r \mp (N+\alpha\cos\theta)} (\Delta\partial_{r}F+\alpha\sin\theta \, \partial_{\theta}F)\right]\,.
\end{eqnarray}
These equations may look complicated, but one can still find an analytic solution. The following is a solution to \eqref{Eqk2}
\begin{multline} \label{valueZ3}
Z_3 = 1 - \ds\frac{4 q_1q_2}{(m-N)} \ds\frac{1}{r-(N+\alpha\cos\theta)} + \ds\frac{4q_1q_2 (p^2-q^2)}{(m-N)^2}\ds\frac{1}{\Sigma} \\ + \left( \ds\frac{4q_1q_2 (p^2-q^2)}{(m-N)^3} - \ds\frac{2 (q+q_1+q_2) (p-q)}{m-N} + \lambda (m-N)\right) \ds\frac{1}{r+(N+\alpha\cos\theta)} ~, 
\end{multline} 
\begin{equation} \label{valueF}
F = F_{nonhom} + F_{hom}\,,
\end{equation}
where
\begin{multline}
F_{nonhom} = \ds\frac{2 q_1 q_2 (p+q)}{(m-N)^2} \log\left[ \ds\frac{\Delta^{1/2} \sin\theta}{[r-(N+\alpha\cos\theta)]^2} \right] - \ds\frac{2q_1+2q_2+p+q}{r_{+}-r_{-}} [r_{+} \log(r-r_{+}) - r_{-} \log(r-r_{-})]\\ - \ds\frac{p^2-q^2}{r_{+}-r_{-}} \left( \ds\frac{2q_1q_2(p+q)}{(m-N)^3} - \ds\frac{q+2q_1+2q_2}{m-N} \right) \log\left[ \ds\frac{r-r_{+}}{r-r_{-}} \right] - \lambda\,{p+q \over 2} \, \log \left( {\sin\theta \over \Delta^{1/2} } \right) \,,
\end{multline}
and
\begin{multline}
F_{hom} = \gamma \left([r-(N+\alpha\cos\theta)] + {2(m-N) \over r_+-r_-}(r_+ \log(r-r_+)-r_-\log(r-r_-)) -\ds\frac{ p^2 - q^2}{r_{+} - r_{-}} \log{\ds\frac{r-r_{+}}{r-r_{-}}} \right) \\ + \kappa \left( \ds\frac{1}{2}\log(\Delta) -\log(\sin\theta) + \ds\frac{m-N}{r_{+} - r_{-}} \log{\ds\frac{r-r_{+}}{r-r_{-}}}\right) \,.
\end{multline}
The function $F_{hom}$ satisfies the equation
\begin{equation}
\mathcal{D}_{+}F_{hom} = 0 \,. 
\end{equation}
In the expressions above $\lambda$, $\gamma$ and $\kappa$ are three constants. The functions $Z_3$ and $F$ presented above are also solutions to the inhomogeneous Laplace equation for $Z_3$, \eqref{EqZ32}, if one imposes the following relation between the constants 
\begin{eqnarray} \label{valuegamma}
	2N\gamma -\kappa = -{\lambda \over 2} \left({m^2-N^2 \over p-q}-(p+q)\right) - { 2q_1q_2(p+q)(m+N) \over (m-N)^3 } + {2N(q+q_1+q_2) \over m-N } \,.
\end{eqnarray}
We now have to make sure that there are no CTCs in the solution. First we rewrite $k$ as
\begin{eqnarray} \label{k2}
	k = {1 \over \Sigma} \left( (\Sigma\mu-\alpha\nu)(d\tau + P_{\theta}\,d\phi) +\nu (\alpha d\tau + P_r\,d\phi )\right)\,.
\end{eqnarray}
To avoid CTCs, one has to make sure that $\nu$ vanishes for $\theta\rightarrow0,\pi$ and that $\Sigma \mu - \alpha \nu$ vanishes for $r\rightarrow r_+$. Using \eqref{muandnuKN}, these conditions lead to the following constraints
\begin{eqnarray}
	\kappa \!\!&=&\!\! -\lambda \,{p+q \over 2} + {2q_1q_2(p+q) \over (m-N)^2 } \,, \label{valuekappa} \\
	\lambda \!\!&=&\!\!\! { 4N(p-q) \over p^2-q^2 -2(m+N)\,r_+ } \!\left( {2q_1q_2(p+q)(m+N) \over (m-N)^4 } \!-\! {(q_1+q_2)(p^2-q^2) \over (m-N)^3} + {(p-q)\,r_+ \over (m-N)^2} \right) \,. \label{valuelambda}
\end{eqnarray}
These relations, together with \eqref{valuegamma}, allow to solve for the constants $(\lambda,\kappa,\gamma)$ in terms of the parameters of the four-dimensional base. The explicit form of $\mu$ and $\nu$ is
\begin{eqnarray}
	\mu &=& \gamma - {2N\gamma \over r+(N+\alpha\cos\theta)} - {4q_1q_2(p+q) \over (m-N)^2}  {\Delta + \alpha^2 \sin^2\theta \over [r-(N+\alpha\cos\theta)]^2[r-(N+\alpha\cos\theta)]}\\ \nonumber 
		&& - \left(  {2q_1q_2(p+q)\over(m-N)^3}(m^2-N^2+p^2-q^2) -(p^2-q^2)\ds\frac{q+2q_1+2q_2}{m-N} + \lambda \, {p+q\over 2}(m-N) \right) {1\over \Sigma} \\ \nonumber 
	&& + \left( {4q_1q_2(p+q) \over (m-N)^2} - (2q_1+2q_2+p+q) \right){ r \over \Sigma } \,, \\
	\nu &=& \left( \gamma -{ 4q_1q_2(p+q) \over (m-N)^2 }{1 \over r-(N+\alpha\cos\theta)} \right) \alpha \sin^2\theta \,,
\end{eqnarray}
with $\lambda$ given by \eqref{valuelambda} and $\gamma$ by 

\begin{eqnarray}
	\gamma = {q+q_1+q_2 \over m-N } - {2 q_1 q_2(p+q) \over (m-N)^3 } -\lambda {m^2-N^2 \over 4N(p-q) } \,.
\end{eqnarray}
Note that the sign of $\gamma$ and $\mu$ in the Kerr-Newman-NUT solution is different from the one for the Reissner-Nordstr\"om solution due to the different choice of orientation of the four-dimensional base.

The parameters of the five-dimensional solution should be chosen such that there are no global CTCs. This analysis is rather lengthy and unilluminating, but it suffices to say that one can always find a choice (or range) of parameters for which the solution is regular and free of global CTCs. As we will see in the next subsection, this range of parameters is even bigger that one could naively expect, because the four-dimensional metric can change signature while the complete five-dimensional solution remains regular and free of CTCs.

\subsection{Ambipolar solution} \label{ambipolar}

An important observation made in \cite{Bena:2005va,Berglund:2005vb} is that one can construct five-dimensional regular and causal supergravity solutions by using a four-dimensional base that changes signature from $(+,+,+,+)$ to $(-,-,-,-)$, as long as the warp factors in the solution change sign in exactly the same way. In this section we will show that the same type of ``ambipolar" solutions can be constructed out of four-dimensional electrovac solutions that change signature. For this purpose, it is useful to rewrite the four-dimensional metric (\ref{KNTBmetric}) as
\be \label{KNTBmetric2}
ds^2_4 = V^{-1} (d\tau + \tilde{P}_\theta d\phi)^2 +V \Bigl({\Delta_\theta\over \Delta} dr^2 + \Delta_\theta d\theta^2 + \Delta \sin^2\theta d\phi^2\Bigr)\,,
\ee
with
\be
\Delta_\theta= \Delta+\alpha^2 \sin^2\theta\,,\qquad V= {\Sigma\over \Delta_\theta}\,,\qquad \tilde{P}_\theta = P_\theta +V \alpha \sin^2\theta\,.
\ee
Recall that the five-dimensional metric is 
\begin{align}
	ds_5^2 = -Z^{-2} (dt + k)^2 + Z ds_4^2 \,.
\end{align}
It is clear that to avoid CTCs, one has to make sure that $g_{\tau\tau}$, $g_{\phi\phi}$ and $ZV$ remain positive everywhere. In particular, it is {\it not} necessary for the functions $Z_I$ to be positive definite, but we only need $Z_I$ and $V$ to have the same sign throughout the whole solution. As $r\to \infty$, $Z_1\sim Z_2\sim Z_3\sim V\sim1$, so all these functions are positive near spatial infinity. Since $\Delta_\theta$ is always positive, $V$ vanishes only when $\Sigma$ does
\be
 \Sigma ~\equiv~ r^2 -(N + \alpha \cos\theta)^2 = [r -(N + \alpha \cos\theta)][r +(N + \alpha \cos\theta)]\,.
\ee
As $r$ decreases, because we have chosen $N>0$, the first term to possibly vanish is $[r-(N+\alpha\cos\theta)]$. There are now two distinct possibilities. In the first one, the parameters are such that $r_+ > \alpha + N$ and thus $V$ never vanishes. In this case, the analysis is similar to the one done in the previous section for the Reisner-Nordstr\"om case. The second possibility is that $r_+ < \alpha + N$, and then $V$ can change sign. But as $[r-(N+\alpha\cos\theta)]$ vanishes, all the poles of the functions determining the solution blow up and the background seems to be highly pathological. We will show here that it is not the case. For this purpose, it is useful to define
\be
 \eta  ~\equiv~  [r -(N + \alpha \cos\theta)]\,.
\ee
At the $\eta=0$ surface the signature of the four-dimensional part of the metric changes from ($+$,$+$,$+$,$+$) to ($-$,$-$,$-$,$-$), and some of the coefficient diverge. However the five-dimensional metric stays completely regular. Indeed, for $\eta\to 0$, we have
\bea \nonumber
 \Sigma &\approx& 2(N+\alpha \cos\theta)\,\eta + \eta^2 + O(\eta^3)\,, \\ \nonumber
 \Delta_\theta &\approx& \left(-2(m-N)(N+\alpha\cos\theta)+p^2-q^2\right) -2(m-N-\alpha\cos\theta)\,\eta) + O(\eta^2)\,, \\ \nonumber
 Z_1 &\approx& -{2q_2(p+q) \over m-N}\,{1 \over \eta} + 1 + O(\eta)\,, \\ 
 Z_2 &\approx& -{2q_1(p+q) \over m-N}\,{1 \over \eta} + 1 + O(\eta)\,, \\ \nonumber
 Z_3 &\approx& 4q_1q_2\left(-{1 \over m-N} + {p^2-q^2 \over 2(m-N)^2(N+\alpha\cos\theta)} \right) \, {1 \over \eta} \\ \nonumber
 && + 1-{q_1q_2(p^2-q^2)(m-3N-2\alpha\cos\theta) \over (m-N)^3(N+\alpha\cos\theta)^2} -{(p-q)(q+q_1+q_2) \over (m-N)(N+\alpha\cos\theta)} + \lambda {m-N \over 2(N+\alpha\cos\theta)} + O(\eta)\,, \\ \nonumber
 \mu &\approx& 2{q_1q_2(p+q) (2(m-N)(N+\alpha\cos\theta)-p^2+q^2) \over (m-N)^2(N+\alpha\cos\theta)}\,{1 \over \eta^2} \nonumber \\ \nonumber
 &&+ \Bigg( {q_1 q_2(p+q) \over (m-N)^2(N+\alpha\cos\theta)}\left( {(p^2-q^2)(m - 2N - \alpha\cos\theta) \over (m-N)(N+\alpha\cos\theta)} +m-N-2\alpha\cos\theta \right) \\ \nonumber
 && -(q_1+q_2+{p+q\over2}) + (q_1+q_2+{q\over2}){p^2-q^2 \over (m-N)(N+\alpha\cos\theta)}  -\lambda {(m-N)(p+q)\over 4(N+\alpha\cos\theta)}\Bigg) {1\over\eta} + O(1)\,, \\ \nonumber
 \nu &\approx& -{4q_1 q_2(p+q)\over (m-N)^2}\alpha\sin^2\theta \, {1\over \eta} + O(1) \,.
\eea
There are possible divergences coming from the coefficient in front of the three-dimensional metric, $ZV$. But as $\eta \rightarrow 0$ we have $Z \sim \eta^{-1}$, $V \sim \eta$, and thus $Z\,V$ is regular.  
The $d\tau^2$ coefficient appears to be very singular, because $ZV^{-1} \sim \eta^{-2}$. However, $g_{\tau\tau}$ has another contribution coming from the angular momentum $k$
\begin{eqnarray}
	g_{\tau\tau} = {Z \over V} - {\mu^2 \over Z^2} \,,
\end{eqnarray}
One can check that the divergences in $Z^{-2}\mu^2$ exactly cancel both the leading and the subleading divergences of $Z V^{-1}$. This ensures that $g_{\tau\tau}$ has a finite value as $\eta \to 0$. The coefficient of $d\phi^2$ can also diverge as $\eta\to0$
\begin{align}
	g_{\phi\phi} = \left({Z\over V} - {\mu^2 \over Z^2}\right)P_{\theta}^2+ 2P_{\theta}\left(Z \alpha\sin^2\theta -{\mu\nu\over Z^2} \right) + \mathrm{finite~terms}\,.
\end{align}
As we discussed above the first term on the right hand side is finite for $\eta \to 0$. One can then check that the $1/\eta$ divergences of $Z \alpha\sin^2\theta$ and $Z^{-2}\mu\nu$ also exactly cancel, which ensure that $g_{\phi\phi}$ stays finite. Finally, one can also easily verify that the off-diagonal terms $g_{t\tau}$ and $g_{t\phi}$ are finite at $\eta=0$. 

The analysis that we performed so far ensures that the five-dimensional solution is regular near the $\eta=0$ surface despite the fact that the four-dimensional base changes signature and seems to be very pathological. We have not presented a detailed analysis of the conditions imposed by global absence of CTCs. As discussed in the previous subsection one can always find a choice of parameters such that the solution is globally causal.

One could expect to find the same kind of ambipolar solution as $[r+(N+\alpha\cos\theta)]\,\rightarrow 0$. A detailed analysis shows that this is not the case, and therefore we should not allow $[r+(N+\alpha\cos\theta)]$ to change sign. This implies that in the ambipolar solution, we should restrict the range of parameter to 
\begin{eqnarray}
	r_+ > \alpha - N \,.
\end{eqnarray}
To conclude this section we would like to point out that such ambipolar solutions are ubiquitous when one looks for BPS solutions of five-dimensional ungauged supergraviy coupled to vector multiplets \cite{Bena:2005va,Berglund:2005vb,Giusto:2004kj,Saxena:2005uk,Bena:2007ju}. It was also recently shown that one can find non-BPS ambipolar solutions with Ricci flat four-dimensional base \cite{Bena:2009qv}. Our solution is a generalization of the one in \cite{Bena:2009qv} and provides further evidence that ambipolar solutions may not be isolated examples among the non-supersymmetric solutions of five-dimensional supergravity.

\subsection{The asymptotic charges}

In this section we calculate the asymptotic charges of the solution, along the lines of section \ref{sec:chargeRN}. First of all, the magnetic charges are given by the same formulae as for the non rotating case
\begin{eqnarray}
	d_1 &=& 2q_1-p+q \,, \nonumber \\
	d_2 &=& 2q_2-p+q \,, \\ \nonumber
	d_3 &=& p+q \,.  
\end{eqnarray}
We now have to compute the electric charges $Q_I$. They are still given by the general formula \eqref{QI}, which yields
\begin{eqnarray}
	Q_1 &=& -{8 \pi^2 \over \kappa} \Bigl( {2(p+q)q_2 \over m-N} -\gamma (q+q_2) \Bigr) \,, \nonumber \\
	Q_2 &=& -{8 \pi^2 \over \kappa} \Bigl( {2(p+q)q_1 \over m-N} -\gamma (q+q_1) \Bigr) \,, \\ \nonumber
	Q_3 &=& -{8 \pi^2 \over \kappa} \Bigl( {4 q_1 q_2 \over m-N} -\gamma (q_1+q_2+p-q)  \\ \nonumber
	 && +2(q+q_1+q_2) {p-q \over m-N} - {4 q_1 q_2 (p^2-q^2) \over (m-N)^3}-\lambda(m-N) \Bigr) \,.
\end{eqnarray}
As for the non-rotating case, $\mu$ goes to a finite non-zero value, $\gamma$, at infinity. One therefore has to introduce the coordinates ($\hat t , \hat \tau $), given by (\ref{hatcoord}) in order to compute the mass, angular momentum and KK charge of the solution. It is also convenient to use the form (\ref{KNTBmetric2}) for the Kerr-Newman-Taub-Bolt metric and to rewrite the
one-form $k$ as
\be
k = \mu (d\tau + \tilde{P}_\theta d\phi) +\tilde{\nu} d\phi\,,
\ee
with
\be
\tilde{\nu}=\nu-\alpha  {\Sigma\over \Delta_\theta}\sin^2\!\theta \, \mu \,.
\ee
One can now rewrite the five-dimensional metric in a form suitable for Kaluza-Klein reduction along $\hat\tau$
\be
ds_5^2 = {\hat{I}_4\over (Z V)^2}\Bigl(d\hat\tau + \left(\gamma - {\mu V^2\over \hat{I}_4} \right) d\hat{t} + \left( \hat{P}_\theta  - {\hat{\nu} \mu V^2\over \hat{I}_4} \right) d\phi \Bigr)^2 + {Z V\over \hat{I}_4^{1/2}} ds^2_E\,,
\ee
where
\be \label{4Dmetric}
ds^2_E = -\hat{I}_4^{-1/2}(d\hat{t}+\hat{\nu} d\phi)^2 + \hat{I}_4^{1/2} \Bigl({\Delta_\theta\over \Delta} dr^2 + \Delta_\theta d\theta^2 + \Delta \sin^2\theta d\phi^2\Bigr)
\ee
is the four-dimensional Einstein metric and
\bea
\hat{I}_4 =  (1-\gamma^2)^{-1}(Z_1 Z_2 Z_3 V - \mu^2 V^2)\,,\quad \hat{P}_\theta = (1-\gamma^2)^{1/2} \tilde{P}_\theta\,,\quad \hat{\nu}=(1-\gamma^2)^{-1/2} \tilde{\nu} \,.
\eea
From this metric, it is easy to read off the mass, $M$ 
\begin{eqnarray}
	M  &=& {1 \over G_4(1 - \gamma^2)} \Bigg[ {m\over2} -(m-N)\gamma^2 -{ q_1q_2 + pq_1 + pq_2 + {q(p-q)\over 2}  \over m-N}   \nonumber \\
	&& +\gamma \left(q_1+q_2+ {p+q\over 2}\right) +q_1q_2 { p^2-q^2 \over (m-N)^3 }+ {\lambda\over 4}(m-N)\Bigg] \,,
\end{eqnarray}
where we introduce the four-dimensional Newton constant 
\begin{eqnarray}
	G_4 = {G_5 \over (1-\gamma^2)^{1/2}}{\kappa \over 2\pi}\,.
\end{eqnarray}
From \eqref{4Dmetric}, one can also read off the angular momentum of the solution
\begin{eqnarray}
	 J = {\alpha \over G_4(1-\gamma^2)^{1/2}} \left( -{2 q_1 q_2 (p+q) \over (m-N)^2} + \left( q_1 + q_2 + {p+q\over2} \right) -\gamma (m-N) \right) \,.
\end{eqnarray}
We finally need the Kaluza-Klein electric and magnetic charges $Q_e$ and $Q_m$, encoded in the one-form 
\begin{eqnarray}
	A_{KK} = \left(\gamma - {\mu V^2\over \hat{I}_4} \right) d\hat{t} + \left( \hat{P}_\theta  - {\hat{\nu} \mu V^2\over \hat{I}_4} \right) d\phi\,.
\end{eqnarray}
Expanding this one form at spatial infinity one finds
\begin{eqnarray}
	 Q_e &=& {1 \over G_4(1 - \gamma^2)} \Bigg[  -\gamma {m\over2} +\gamma (1+\gamma^2){N\over2} - \gamma { q_1q_2 + pq_1 + pq_2 + {q(p-q)\over 2} \over m-N}   \nonumber \\
	&& +{1+\gamma^2\over2} \left(q_1+q_2+ {p+q\over 2}\right) +\gamma q_1q_2 { p^2-q^2 \over (m-N)^3 }+ \gamma {\lambda \over 4}(m-N)\Bigg] \,,
\end{eqnarray}
and
\begin{eqnarray}
	 Q_m = -(1-\gamma^2)^{1/2} {N \over 2G_4} \,.
\end{eqnarray}
Finally, one can compute the rest mass of the solution
\be
	M_0 \equiv (1-\gamma^2)^{-1/2}(M-\gamma Q_e) = {\pi\over G_5 \kappa} (m+N) + Q_m + {1 \over 16 \pi G_5}\left(Q_1+Q_2+Q_3\right)\,.
\label{massKNTB}
\ee
It is clear from this expression that the solution has the same mass and charges as a non-extremal black hole. The mass of the five-dimensional solution is a sum of the electric charges and the solitonic charges of the four-dimensional base. The dependence of the mass on the charges is still linear due to the ``floating brane" Ansatz.

\section{Conclusions}

Starting from a four-dimensional Euclidean background that solves Einstein-Maxwell equations, we found a six-parameter family of solutions to five-dimensional $\mathcal{N}=2$ ungauged supergravity coupled to two vector multiplets. Our solutions are regular, horizonless, do not preserve any supersymmetries and have the same charges at infinity as a non-extremal black hole. They generalize substantially the solutions found in \cite{Bena:2009qv} which were based on a Ricci-flat four-dimensional base and had only self-dual (or anti-self-dual) fluxes. The key point of the construction, in both \cite{Bena:2009qv} and our work, is the existence of a bolt in the four-dimensional base \cite{Gibbons:1979xm}, on which one can put magnetic fluxes. These fluxes provide non-singular sources for the warp factors of the solution, ensure its regularity and are ultimately responsible for the charges at spatial infinity. It would be interesting to construct other non-supersymmetric five-dimensional supergravity solutions with a four-dimensional electrovac base. If this base space has interesting topology one should be able to find regular solutions by putting fluxes on it. There are some well-known backgrounds that could be used for such a construction. The ten-parameter family of solutions constructed by Carter \cite{Carter:1968ks} is a notable example, which includes the Kerr-Newman solution. Another interesting example is the Euclidean Melvin solution \cite{Melvin:1963qx}. This solution is not asymptotically flat and may lead to non-supersymmetri solutions with interesting asymptotic structure. Trying to build five-dimensional solutions on these spaces may be challenging, but the presence of two commuting Killing vectors on the four-dimensional base should render the problem tractable.

In the supergravity action \eqref{5daction}, the three gauge fields have symmetric roles. This symmetry is explicitly broken by our assumptions \eqref{assumptions}, which leads to a linear system of differrential equations. A very natural question is whether one can put all three $U(1)$ gauge fields on the same footing, and find solutions which are symmetric under the interchange of the three gauge fields. While the ``floating brane" Ansatz presumably allows for such solution, it seems to be a rather difficult task to find completely general solutions in this Ansatz. Indeed, turning on $\omega_-^{(1)}$ and $\omega_-^{(2)}$ modifies the equations of motion and they can no longer be solved in a linear way.

As we discussed above the solutions constructed in this paper can be obtained by compactifying eleven-dimensional supergravity on $T^6$ with three sets of M2 and M5 branes wrapping two- and four-cycles on the torus. It should be in principle straightforward to construct analogous compactifications replacing the $T^6$ by an arbitrary Calabi-Yau threefold. These would correspond to solutions of five-dimensional $\mathcal{N}=2$ ungauged supergravity coupled to $h_{1,1}-1$ vector multiplets, where $h_{1,1}$ is one of the Hodge numbers of the Calabi-Yau. In the BPS case such solutions were discussed in \cite{Cheng:2006yq}. 

Rather than finding new solutions by solving the equations of motion, a very fruitfull approach is the use of solution generating techniques. In this context, it is useful to note that the solutions discussed in this paper have at least two commuting space-like Killing vectors. This symmetry can be utilized to generate an even more general class of non-extremal solutions by using spectral flow \cite{Bena:2008wt}. This may proceed in the following way - first one has to use the results of \cite{Bena:2008dw} to dualize the eleven-dimensional solution to IIB supergravity and then perform the spectral flow transformation of \cite{Bena:2008wt}. The action of spectral flow on non-BPS supergravity solutions has already shown its efficiency \cite{Bena:2009fi}, \cite{AlAlawi:2009qe}, and it is natural to expect that it will be useful for generating new interesting solutions.

The construction of our solutions relies on the ``floating brane" Ansatz of \cite{Bena:2009fi}, which states that the metric warp factors and the electric potentials are related. All the solutions found so far within this Ansatz have a mass that is linear in the sum of the electric charges. It should be expected that for a generic non-supersymmetric supergravity solution this linear dependence should not be present. Very few such more general non-BPS solutions are known \cite{Jejjala:2005yu,Elvang:2004xi,RL} and it would be very interesting to find more of them. It is also worth exploring the limitations on the types of solutions that can be constructed via the ``floating brane" Ansatz and to find new more general techniques for constructing non-BPS solutions. 

An interesting open question is whether the solutions constructed in this paper are stable. Since the solutions have the same asymptotics as a non-exrtremal black hole, one can expect that they will be unstable, it will be very interesting to understand the details of this putative instability. We have not performed the stability analysis of our solutions and we expect this to be a non-trivial task, see \cite{Gross:1982cv} for a discussion of the instability of the Schwarzschild instanton. It is known that the regular non-BPS solutions found in \cite{Jejjala:2005yu} are unstable \cite{Cardoso:2005gj}. It was later shown that this instability has a natural interpretation in terms of Hawking radiation \cite{Chowdhury:2007jx}. It is tempting to speculate that if the non-BPS solutions presented here are unstable their instability should also be interpreted as Hawking radiation for the corresponding non-extremal black hole with the same asymptotic charges.

\bigskip
\leftline{\bf Acknowledgements}
\smallskip

We would like to thank Iosif Bena, Stefano Giusto and Nick Warner for numerous helpful discussions. The work of NB was supported
in part by DOE grant DE-FG03-84ER-40168. The work of CR was supported in part by the DSM CEA-Saclay, by the ANR grant BLAN06-3-137168, and by the Marie Curie IRG String-QCD-BH. NB would like to thank the IPhT, CEA Saclay for hospitality while this work was initiated. CR would like to thank the Physics and Astronomy Department at USC for hospitality.

\section*{Appendix A. Extremal Reissner-Nordstr\"om}
\appendix
\renewcommand{\theequation}{A.\arabic{equation}}
\setcounter{equation}{0} 

An interesting limiting case of the solution presented in Section 3 is when the two horizons of the four-dimensional base coincide. This is the extremal Euclidean dyonic Reissner-Nordstr\"{o}m background
\begin{equation}
ds^2_{4} = \left(1 - \frac{m}{r}\right)^2 d\tau^2 + \left(1 - \frac{m}{r}\right)^{-2} dr^2 + r^2 (d\theta^2+\sin^2\theta d\phi^2)~,
\end{equation}
\begin{equation}
F = \ds\frac{2q}{r^2} \, d\tau \wedge dr + 2 p \sin\theta \,d\theta \wedge d\phi \,.
\end{equation}
This background is a limit of the dyonic Reissner-Nordstr\"{o}m black hole which is obtained by taking $m^2=p^2-q^2$. The two horizons degenerate and we have 
\begin{equation}
r_{+} = r_{-} = m \,.
\end{equation}
The near horizon limit of the Lorentzian extremal Reissner-Nordstr\"om black hole is the Bertoti-Robinson solution which is $AdS_2\times S^2$ with electric and magnetic flux \cite{Bertotti:1959pf}. In the Euclidean solution of interest the horizon has become a bolt of radius $m$ and near the bolt we can set
\begin{equation}
r = m +  \ds\frac{m^2}{\rho^2} \,,
\end{equation}
and rewrite the metric as
\begin{equation}
ds^2_{NH} = m^2 \left( \ds\frac{d\rho^2 + d\tau^2}{\rho^2} +  d\theta^2+\sin^2\theta d\phi^2 \right)\,.
\end{equation}
This is the metric on $H_2^{+}\times S^2$, where $H_2^{+}$ is the Poincar\'e half plane and we have the following range of coordinates $\tau \in (-\infty,\infty)$ and $\rho \in (0,\infty)$. Note that we still have a finite size bolt ($S^2$) at $r=m$ on which we can put flux. At asymptotic infinity the metric approaches the flat metric on $\mathbb{R}^4$. This should be contrasted with the case of the non-extremal Euclidean Reissner-Nordstr\"om black hole of Section 3, where we had to periodically identify the coordinate $\tau$ to get a regular metric near the outer horizon. The five-dimensional supegravity solution based on this four-dimensional base has the same warp factors and fluxes as the solution in Section 3, however one should remember to set $m^2=p^2-q^2$. The coordinate $\tau$ is non-compact but it is still an isometry of the five-dimensional solution. This means that we have the electric charges corresponding to the three $U(1)$ gauge fields smeared along $\tau$. What happens effectively is that in the extremal limit the coordinate $\tau$ decompactifies and the five-dimensional solution is asymptotic to $\mathbb{R}^{1,4}$ and corresponds to a smeared distribution of charges along $\tau$. With this in mind one can proceed in the same way as in Section 3 and compute the asymptotic charges and mass densities of the five-dimensional solution\footnote{Note that, since the $\tau$ coordinate is not compact anymore, we are now computing charge and mass densities.} 
\bea 
 Q_1 \!\!\!&=&\!\!\! - 4\pi \left(  {2(p+q)q_2 \over  m} + \gamma (q+q_2)\right)\,, \notag\\
 Q_2 \!\!\!&=&\!\!\! - 4\pi \left(  {2(p+q)q_1 \over  m} + \gamma (q+q_1)\right)\,, \\
 Q_3 \!\!\!&=&\!\!\! - 4\pi \left(  {4 q_1 q_2 \over  m} + \gamma (q_1+q_2+p-q) + {2(p-q)(q+q_1+q_2) \over m} - {4q_1q_2(p^2 - q^2) \over m^3} \right)\,, \notag\\
 M_0 \!\!\!&=&\!\!\!  \ds\frac{1}{16\pi G_5} \left(8\pi m +Q_1+Q_2+Q_3 \right)~. \notag
\eea
It is clear from the dependence of the mass on the charges that we again have a non-BPS five-dimensional solution that has the same asymptotic charges as a non-extremal black hole. This may seem somewhat strange because we have started with an extremal four-dimensional solution, which is also known to be BPS\footnote{The Lorentzian extremal Reissner-Nordstr\"om solution is a BPS background interpolating between $AdS_2\times S^2$ and $\mathbb{R}^{1,3}$. Going to the Euclidean regime does not spoil the supersymmetry of the solution which now interpolates between $H_2^{+}\times S^2$ and $\mathbb{R}^4$, see for example \cite{Lu:1998nu}.}. There is nothing puzzling going on here, to get the five-dimensional solution we have added fluxes to the four-dimensional base which break the supersymmetry completely. In addition the difference between the mass and the sum of the electric charges corresponds to the ``solitonic'' contribution of the bolt, and therefore one should not expect to have a solution with the same charges as an extremal black hole. 

Finally we will provide some comments on the extremal limit of the Kerr-Newmann-NUT solution of Section 4. This limit arises when we have $m^2 = N^2 - \alpha^2 + p^2 - q^2$. There is no need to compactify the coordinate $\tau$ near $r=r_{+}=m$, however we still have to compactify $\tau$ to ensure regularity at spatial infinity. Since we will have an unique identification of $\tau$, $\tau \sim \tau + 8\pi N$, we will not have to impose the constraint \eqref{paramrelation}. The five-dimensional solution will be the same as in Section 4 and will still be asymptotic to $\mathbb{R}^{1,3}\times S^1$.

The extremal Lorentzian Kerr background was discussed in \cite{Bardeen:1999px} and has been given a holographic interpretation in \cite{Guica:2008mu}. This has also been generalized to the extremal Kerr-Newman solution \cite{Hartman:2008pb}. Note that these papers consider exclusively the Lorentzian backgrounds. For $N=p=q=0$ there seems to be no good analytic continuation of the Kerr metric to Euclidean space. However when at least one of the parameters $(N,q,p)$ is not zero we can have an extremal Euclidean background, as discussed above. It will be very interesting to see if one can use the extremal Euclidean Kerr-Newman-NUT background to construct a five-dimensional solution with a scaling region and asymptotic symmetries that realize a copy of the Virasoro algebra.



\end{document}